\documentclass{article}
\usepackage{spconf,amsmath,graphicx,hyperref}
\usepackage{sidecap}
\usepackage{algorithm}
\usepackage{bbm}
\usepackage{booktabs}
\usepackage{multirow}
\usepackage{algorithmicx}
\usepackage{algpseudocode}
\usepackage{amsfonts}
\usepackage{graphicx}  
\usepackage[dvipsnames]{xcolor}

\def\ie{{\textit{i.e., }}}
\def\eg{{\textit{e.g., }}}

\title{AR\&D: A Framework for Retrieving and Describing Concepts for Interpreting AudioLLMs}
\name{Townim Faisal$^{\star,\dagger}$ \qquad Ta Duc Huy$^{\star}$ \qquad Siqi Pan$^{\dagger}$ \qquad Jeremy Stoddard$^{\dagger}$ \qquad Zhibin Liao$^{\star,\ddagger}$}
\address{
$^{\star}$ Australian Institute for Machine Learning, University of Adelaide, 
$^{\dagger}$ Dolby Laboratories, \\
$^{\ddagger}$ School of Computer and Mathematical Sciences, University of Adelaide, Australia
\thanks{Work done during the Research Internship at Dolby Laboratories.}
}

\begin{document}
\maketitle
\begin{abstract}
Despite strong performance in audio perception tasks, large audio-language models (AudioLLMs) remain opaque to interpretation.
A major factor behind this lack of interpretability is that individual neurons in these models frequently activate in response to several unrelated concepts.
We introduce the first mechanistic interpretability framework for AudioLLMs, leveraging sparse autoencoders (SAEs) to disentangle polysemantic activations into monosemantic features. Our pipeline identifies representative audio clips, assigns meaningful names via automated captioning, and validates concepts through human evaluation and steering. Experiments show that AudioLLMs encode structured and interpretable features, enhancing transparency and control. This work provides a foundation for trustworthy deployment in high-stakes domains and enables future extensions to larger models, multilingual audio, and more fine-grained paralinguistic features. 
\end{abstract}


\section{Introduction}
Large audio-language models (AudioLLMs) have recently demonstrated impressive capabilities in tasks such as audio classification, emotion recognition, captioning, and environmental sound understanding~\cite{chu2024qwen2,wang2024audiobench,tangsalmonn}. These models learn rich, high-dimensional representations that encode complex acoustic patterns, enabling state-of-the-art performance across a wide range of applications. However, despite their success, the internal mechanisms by which these models encode and represent audio information remain largely opaque. In particular, individual neurons often exhibit \textit{polysemantic} behavior~\cite{elhage2022toy}, responding to multiple unrelated audio concepts. Such behavior limits the interpretability of AudioLLMs, making it challenging to understand model decisions, validate learned concepts, or deploy these models in high-stakes domains, such as clinical audio analysis or assistive technologies~\cite{jiao2024audio,CarreiroMartins2024}.

Mechanistic interpretability techniques, such as circuit analysis~\cite{olsson2022context} and sparse autoencoders (SAEs) ~\cite{huben2023sparse}, have been widely used to understand large language models (LLMs) by identifying neurons or circuits corresponding to specific concepts. While some methods have been extended to large multimodal models (LMMs), their use is largely limited to vision~\cite{zhang2024largemultimodalmodelsinterpret,lou2025sae}, leaving audio models underexplored. Another challenge is, unlike text-only LLMs, AudioLLMs encode paralinguistic cues (\eg tone, emotion) that can change the interpretation of identical words~\cite{kang2024frozen,kim2024paralinguistics}. For example, the phrase \textit{``I'm fine"} may indicate well-being when spoken cheerfully but convey frustration or anger when spoken angrily, a distinction that text-only LLMs cannot detect. However, current interpretability methods neither account for these cues nor reveal which latent features represent them.
Here, a latent feature is referred to as a single neuron's output, equivalent to an element in a latent feature vector produced by an intermediate model layer.

\begin{figure*}[!htbp]
    \centering
    \includegraphics[width=1.0\linewidth]{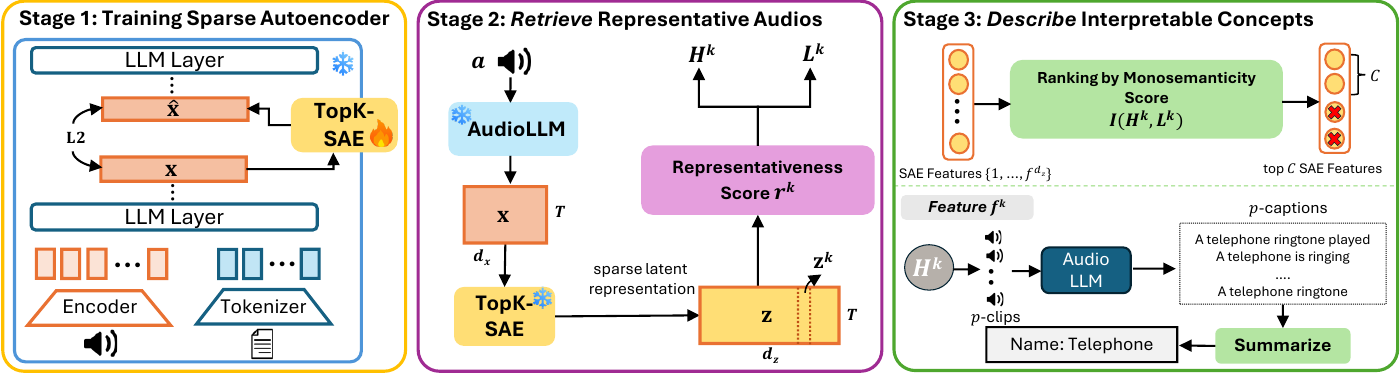}
    \vspace{-1em}
    \caption{Overview of the \textbf{Audio Retrieve and Describe (AR\&D)} pipeline to discover and naming interpretable concepts in AudioLLM. \textit{Stage 1}: The SAE is trained to reconstruct representations $\mathbf{x}$ from the AudioLLM, yielding a latent space of sparse, monosemantic features. \textit{Stage 2}: Using a probing dataset $\mathcal{A}$, we compute SAE activations $\mathbf{Z}$ and calculate a representativeness scores by $F(\cdot)$ for each feature, selecting the $p$ most and least representative audio clips ($H^k$ and $L^k$) per feature. \textit{Stage 3}: We filter top features using monosemanticity scores derived from $H^k$ and $L^k$, and interpret them by generating and summarizing captions from representative clips $H^k$, producing a final set of human-understandable concepts.}
    \label{fig:pipeline}
\end{figure*}

To address these challenges, we propose \textbf{Audio Retrieve and Describe (AR\&D)}, the first mechanistic interpretability pipeline (shown in Fig.~\ref{fig:pipeline}) tailored for AudioLLMs that extends sparse autoencoder (SAE) methods. 
AR\&D introduces three components. 
\textit{(i)} Disentangling polysemantic features into monosemantic ones~\cite{elad2010sparse,olshausen1996emergence}. 
Here, a monosemantic feature consistently captures a single characteristic (\eg rising pitch), and a concept is a feature with a human-understandable label (\eg `high tone'). 
\textit{(ii)} Introducing a representative score to automatically identify audio clips that best characterize each feature. 
\textit{(iii)} Filtering the most monosemantic features and uses AudioLLM-generated captions to assign interpretable, human-readable concept names.
We further demonstrate that these concepts are meaningful through human evaluation and feature steering, confirming their alignment with human perception and causal influence on model behavior.

\section{Audio Retrieve and Describe (AR\&D)}

\subsection{Training Sparse Autoencoder (Stage 1)}
Sparse Autoencoders (SAEs) have been used to unfold polysemantic features into monosemantic and interpretable features in LLMs~\cite{huben2023sparse, dreyer2025mechanistic}. 
Let $\mathbf{x} \in \mathbb{R}^{T \times d_x}$ be the output of a LLM layer $l$ in an AudioLLM, where $T$ is the number of tokens and $d_x$ the hidden dimension. 
We apply a TopK Sparse Autoencoder (TopK-SAE)~\cite{makhzani2013k,gaoscaling} to reconstruct $\mathbf{x}$ from a sparse latent representation:
\begin{align}
\mathbf{z}  = \mathrm{TopK}(\mathbf{W}_{\text{enc}}\mathbf{x} + \mathbf{b}_\text{enc}), 
\quad
\mathbf{\hat{x}}  = \mathbf{W}_{\text{dec}} \mathbf{z} %
\label{eq:sae}
\end{align}
where $\{\mathbf{W}_{\text{enc}}$,$\mathbf{W}_{\text{dec}}, \mathbf{b}_\text{enc}\}$ are the learnable parameters of the TopK-SAE.
The sparse latent dimension $d_z = e \times d_x$.
Here $e$ is the expansion factor which allows a larger and more expressive feature space, but only the $K$ most active units per token are retained by the $\mathrm{TopK}$ operator, enforcing sparsity. The parameters are trained with an L2 reconstruction loss: $|\mathbf{x} - \mathbf{\hat{x}}|^2_2$.

\subsection{\textit{Retrieve} Representative Audio for SAE Features (Stage 2)}
To analyze the semantics of each latent feature $f^k$ \ie the $k$-th dimension of the SAE output $\mathbf{z}$, we first identify the \emph{most} and \emph{least} representative audio clips of $f^k$. Given a probing dataset $\mathcal{A} = \{a_i\}_{i=1}^N$ of $N$ audio clips $a_i$.
Each audio clip (noted as $a$ without subscript $i$ for simplicity) is encoded by an AudioLLM to yield the representation $\mathbf{x}$, which then passed to TopK-SAE to produce the latent $\mathbf{z} = \{\mathbf{z}^k\}_{k=1}^{d_z} \in \mathbb{R}^{T \times d_z}$, with $T$ denoting the number of tokens in $a$ and $\mathbf{z}^k \in \mathbb{R}^{T}$ denoting the activation values in the $k$-th feature $f^k$. 
Note that $T$ can be different for different clips. 
The \textit{representativeness score} $r^k$ for $f^k$ is defined as the product of two measures: the \textit{mean activation} $\mathbf{\mu}^k = \frac{1}{T} \sum_{t=1}^{T} \mathbf{z}_{t}^k$, which quantifies the average activation strength of the feature across the tokens, and the \textit{coverage} $c^k = \frac{1}{T} \sum_{t=1}^{T} \mathbf{1}\left[ \mathbf{z}_{t}^k > 0 \right]$, which reflects the proportion of tokens for which the feature is active. Formally, we define a scoring function $F:\mathbb{R}^{T}\to\mathbb{R}$ that maps the SAE activation sequence $\mathbf{z}^k$ to a scalar value:
\begin{equation}
\label{eq:representative}
   r^k = F(\mathbf{z}^k) = \mu^k \cdot c^k, 
\end{equation}
which summarizes both the magnitude and temporal prevalence of $f^k$.
Prior work~\cite{zhang2024largemultimodalmodelsinterpret} selects representative image patches using only mean activation. While this is reasonable for images, where neighboring pixels often share semantic content, it is insufficient for audio: individual audio tokens are short and may not capture complete or meaningful sounds. Consequently, high mean activation over a brief segment does not guarantee that the feature is consistently expressed, and coverage alone may emphasize weak but temporally extended activations. By combining mean activation and coverage, our representativeness score ensures that selected clips reflect both the strength and temporal extent of a latent feature $f^k$, providing a more reliable characterization of audio features as shown in Table~\ref{tab:interpretability}.

We then compute the representativeness score for each feature $f^k$ across all audio clips in $\mathcal{A}$, obtaining one $r_i^k$ score for each $a_i$. 
The audio clips are ranked by their respective $r_i^k$ scores.
Then the top $p$ audio clips with the highest $r_i^k$ scores are selected as the most representative examples, denoted as $H^k$. Similarly, the lowest $p$ clips, noted as $L^k$, are selected as the least representative examples, meaning the feature is minimally or not expressed in those clips. %
\begin{figure}[!tbp]
    \centering
    \includegraphics[width=1.0\linewidth]{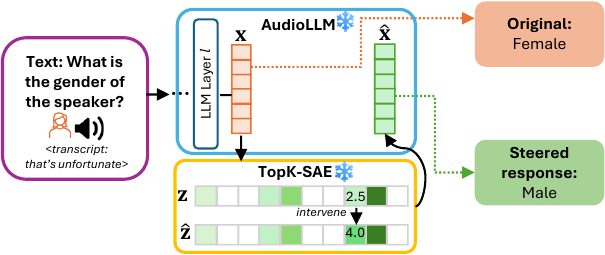}
    \vspace{-1em}
    \caption{Illustration of the steering mechanism. From a given layer of AudioLLM, the obtained input $\mathbf{x}$ is transformed into an SAE representation $\mathbf{z}$. The $k$-th feature (\ie the targeted concept) is then replaced with a predefined steering value (\eg up from 2.5 to 4.0 in the example). The modified representation $\hat{\mathbf{z}}$ is processed by a TopK operator and decoded into $\hat{\mathbf{x}}$ (Eq.~\ref{eq:sae}), which is subsequently fed through the rest AudioLLM layers (as marked by the dotted lines), replacing $\mathbf{x}$, allowing fine-grained control over specific features in the model.}
    \label{fig:steering}
\end{figure}
\subsection{\textit{Describe} Interpretable Concepts (Stage 3)}
\textbf{Interpretable Feature Identification.} 
Features that consistently capture a single, coherent concept are easier to interpret. To identify such features, we quantify their \textit{monosemanticity}~\cite{pach2025sparse}, measuring how selectively and coherently a feature responds to its most representative audio clips versus least. For each feature $f^k$, we consider its the most/least representative sets $H^k$ and $L^k$, embedding each sample $a$ with CLAP~\cite{elizalde2023clap} to obtain a semantic embedding $\mathbf{e}$. 
The intra-set coherence (within $H^k$ or $L^k$) is measured as the average pairwise cosine similarity:
\begin{equation}
E_{G}=\frac{1}{B}\sum_{\substack{\mathbf{e}_i,\mathbf{e}_j\in G, \mathbf{e}_i \neq \mathbf{e}_j}}\mathrm{sim}(\mathbf{e}_i,\mathbf{e}_j), G\in\{H^k,L^k\},
\end{equation}
where $B = \binom p2$ denotes the normalisation factor that counts the number of evaluated cosine similarities (sim).
The monosemanticity score is defined as: 
\begin{equation}
m^k= I(H^k, L^k) = \frac{E_{H^k} - E_{L^k}}{\sigma_\text{pooled}+\varepsilon},
\end{equation}
where $\sigma_\text{pooled} = \sqrt{\frac{\sigma_{H^k}^2 + \sigma_{L^k}^2}{2}}$ normalizes for variability across both sets, and $\varepsilon$ ensures numerical stability. Intuitively, $m^k$ favors features whose most representative clips form tight, coherent clusters while the least representative remain dispersed. A large positive value of $m^k$ indicates a feature that consistently activates for a semantically homogeneous concept and discriminates from counterexamples. 
Collecting all scores $\{m^k\}_{k=1}^{d_z}$, we rank and select the top-$C$ $f^k$ features. 
Restricting analysis to the top-$C$ ensures focus on features that are both \textit{coherent} and \textit{distinctive}, filtering out noisy or ambiguous ones.

\noindent\textbf{Automatic Explaining of Interpretable Features.} 
For each selected top-$C$ features $f^k$, we obtain the $p$ most representative clips from its $H^k$ and query an AudioLLMs (\ie we use SeaLLM-Audio-7B~\cite{SeaLLMs-Audio} to generate captions) with the prompt: ``\textit{Generate a detailed caption for the audio clip}''. The resulting responses are aggregated and provided to the Llama-3-70B-Instruct model with the prompt: ``\textit{Describe the common sound-related concept present among these captions}''. This procedure yields concise, semantically consistent explanation that capture the shared acoustic properties of the most highly activated audio segments.

\section{Experiments}
\textbf{Implementation Details.} 
We train TopK-SAE~\cite{gaoscaling} on train set of WavCaps~\cite{mei2024wavcaps} (108317 clips) and IEMOCAP~\cite{busso2008iemocap} (10039 clips) to interpret the pretrained Qwen2-Audio-7B-Instruct model~\cite{chu2024qwen2}. We use both of these datasets as our probing dataset to find the representative audio clips. Activation vectors are pre-extracted from the residual stream after layers $l \in \{5,16,26\}$. Each SAE is constrained to at most $K=250$ non-zero latent units and trained with expansion factors $e \in \{4, 8, 16\}$. Models are optimized for $10^5$ steps using Adam optimizer with a batch size of 4096 and learning rate $1\mathrm{e}{-5}$. Unless otherwise noted (ablation studies), results are reported for $l=26$ and $e=8$. Empirically, we found that $p=4$ and $C=5000$ are sufficient to discover and name interpretable concepts. We provide a selection of identified concepts for each layer, along with representative audio clips and steering examples\footnote{\url{https://bit.ly/autointerpret-audiollm}}.

\noindent\textbf{Compared Methods.} 
We compare AR\&D against four methods: 
\textit{(a) Polysemantic Features (Poly. Feats.)}, which directly uses the original activations $\mathbf{x}$ from the target model~\cite{chu2024qwen2} without applying the SAE, \ie no sparse latent representations are used in the pipeline, 
\textit{(b) Random Representatives (Rand. Rep.)}, which samples random $p$ audio clips as representatives for each $f^k$ in the top-$C$ monosemantic features and averages over 10 runs, \textit{(c) Mean Activation (Mean Act.)}, which adapts AutoInterpret-V~\cite{zhang2024largemultimodalmodelsinterpret} by using only the mean activation ($\mu^k$) in Eq.~\ref{eq:representative} as the representative score within our pipeline, and \textit{(d) Coverage}, which uses only coverage ($c^k$) in Eq.~\ref{eq:representative} as the representative score. All of the compared methods have been experimented on the 26th layer of the target model where SAE's expansion factor is 8.

\section{Results and Discussion}
\textbf{Interpretability Evaluation.} 
We evaluate our method on the FSD50k~\cite{fonseca2021fsd50k} test set (10231 samples), with results shown in Table~\ref{tab:interpretability}. To measure semantic alignment between predicted and reference concepts, we use CLAP~\cite{elizalde2023clap}. Precision, recall, and F1 are computed by matching predictions to references via the Hungarian algorithm~\cite{kuhn1955hungarian}, considering only matches above a threshold $\gamma = 0.7$, so that each prediction is assigned to at most one reference. We also report mean Average Precision (mAP) by ranking all prediction–reference pairs by similarity and integrating precision across all ranks. Additionally, we report the highest monosemanticity scores (MS) to indicate the interpretability of the most coherent features. 
Our observations are as follows: 
\textit{(1)} Our method consistently achieves the best results across all metrics. Compared to the second-best method (Coverage), it improves F1 by 33\% and mAP by 49\%, demonstrating the effectiveness of disentangled features and our scoring formulation in aligning predicted concepts with reference semantics. 
\textit{(2)} `Rand. Rep.' performs slightly better than `Poly. Feats.', suggesting that averaging over randomly selected clips for monosemantic features provides a weak stabilizing effect that reduces noise and yields higher semantic alignment than raw, polysemantic features. 
\textit{(3)} Both methods (Mean Act. and Coverage) achieve higher performance than `Poly. Feats.' and `Rand. Rep.', highlighting the advantages of our representative score mechanism over random selection or direct use of activations.
\begin{table}[!tbp]
\centering
\resizebox{1.0\columnwidth}{!}{
\begin{tabular}{c|c|c|c|c|c}
\toprule
Method & MS $\uparrow$ & Precision $\uparrow$ & Recall $\uparrow$ & F1 $\uparrow$ & mAP $\uparrow$ \\
\midrule
Poly. Feats.~\cite{chu2024qwen2} & 1.14 & 0.05 & 0.20 & 0.08 & 0.05 \\
Rand. Rep. & 2.13 $_{\pm 0.08}$ & 0.11 $_{\pm 0.05}$ & 0.23 $_{\pm 0.06}$ & 0.15 $_{\pm 0.04}$ & 0.10 $_{\pm 0.05}$ \\
Mean Act.~\cite{zhang2024largemultimodalmodelsinterpret} & 4.76 & 0.42 & 0.42 & 0.42 & 0.36 \\
Coverage & 5.28 & 0.46 & 0.44 & 0.45 & 0.39 \\
\textbf{AR\&D} ($l$=5) & 9.17 & 0.48 & 0.51 & 0.49 & 0.47 \\
\textbf{AR\&D} ($l$=16) & 9.31 & 0.52 & 0.58 & 0.55 & 0.53 \\
\textbf{AR\&D} ($l$=26) & \bf 9.33 & \bf 0.55 & \bf 0.65 & \bf 0.60 & \bf 0.58 \\
\bottomrule
\end{tabular}
}
\caption{Evaluation of interpretability on the FSD50k dataset, showing highest monosemanticity scores (MS), precision, recall, F1, and mean Average Precision (mAP). Higher values indicate better semantic alignment and feature interpretability.}
\label{tab:interpretability}
\end{table}

\noindent\textbf{Human Evaluation.} 
To assess the reliability of our automatic concept naming framework, we conducted a human study on 50 randomly selected concepts. Expert annotators listened to the most representative audio clips for each concept, provided descriptive labels, and rated how well the automatically assigned names reflected the core acoustic properties on a 0–5 scale. Table~\ref{tab:human} reports both the expert-rated scores and semantic similarity scores (\ie CLAP cosine similarity and BERTScore~\cite{bert-score}) between generated names and expert descriptions. Our method significantly outperforms the baseline approach, achieving higher alignment with human judgment and stronger semantic agreement across both metrics.
\begin{table}[!htbp]
\centering
\resizebox{0.95\columnwidth}{!}{
\begin{tabular}{c|c|c|c}
\toprule
Method & Expert-rated score $\uparrow$ & CLAP Cosine $\uparrow$ & $\text{BERTScore} \uparrow$  \\
\midrule
Poly. Feats. & 2.13 $_{\pm 0.79}$ & 0.47 $_{\pm 0.41}$ & 0.23 $_{\pm 0.41}$ \\
\textbf{AR\&D} & \bf 4.29 $_{\pm 0.81}$ & \bf 0.84 $_{\pm 0.20}$ & \bf 0.92 $_{\pm 0.11}$ \\
\bottomrule
\end{tabular}
}
\caption{Human and semantic similarity evaluations.}
\label{tab:human}
\end{table}

\noindent\textbf{Steering.} 
We follow the steering procedure from~\cite{zhang2024largemultimodalmodelsinterpret} (shown in Fig.~\ref{fig:steering}) and report \textit{sensitivity} in Table~\ref{fig:steering}, defined as the percentage of samples whose predictions shift to the target concept after steering. Experiments are conducted on IEMOCAP-Emotion (1004 samples) and VoxCeleb1-Gender (4874 samples) from AudioBench~\cite{wang2024audiobench}, using Llama-3-70B-Instruct as a judge model. We map features semantically linked to emotion or gender into target categories (Happy, Angry, Female). `Poly. Feats.' shows very low sensitivity, indicating that raw polysemantic activations are difficult to control. In contrast, our method achieves consistently high sensitivity across all tasks, demonstrating effective and interpretable feature steering.
The lower score for \textit{Happy$\rightarrow$Angry} likely reflects the limited number of `Angry' samples, which reduces steering reliability in this case.
\begin{table}[!tbp]
\centering
\resizebox{1.0\columnwidth}{!}{
\begin{tabular}{c|c|c|c|c}
\toprule
Method &  Neutral $\rightarrow$ Happy & Sad $\rightarrow$ Happy & Happy $\rightarrow$ Angry & Female $\rightarrow$ Male  \\
\midrule
Poly. Feats. & 0.13 & 0.08 & 0.04  & 0.09\\
\textbf{AR\&D} &  \bf 0.75 & \bf 0.68  & \bf 0.47 & \bf 0.61\\
\bottomrule
\end{tabular}
}
\caption{Evaluation of feature steering \textit{sensitivity} on targeted emotion and gender concepts. Higher scores indicate a stronger effect of the intervention on the model’s output.}
\label{tab:steering}
\end{table}

\noindent\textbf{Ablation Studies.} 
We conducted an ablation study on three different layers $l$ of the AudioLLM model and report the highest monosemanticity scores in Table~\ref{tab:ms}. The `Poly. Feats.' setting shows consistently low scores across all layers, reflecting the high polysemanticity of raw activations. In contrast, AR\&D substantially improves monosemanticity at each layer, demonstrating its effectiveness in disentangling polysemantic features. Notably, deeper layers (layer 26) achieve the highest scores, indicating that semantic disentanglement is more pronounced in deep layers. 
Additionally, increasing the expansion factor beyond 8 yields only minor gains while increasing significant computational cost, suggesting moderate expansion is sufficient.
\begin{table}[!hbp]
\centering
\resizebox{0.65\columnwidth}{!}{
\begin{tabular}{c|c|c|c|c}
\toprule
\multirow{2}{*}{Method} & Expansion &  \multicolumn{3}{c}{Layer}\\
\cline{3-5}
& Factor & 5 & 16 & 26 \\
\midrule
Poly. Feats. & - & 0.68 & 1.02 & \bf 1.14\\
\midrule
\multirow{3}{*}{\textbf{AR\&D}} & 4 & 6.86 & 7.34 & \bf 9.17\\
& 8 & 7.01 & 7.85 & \bf 9.31\\
& 16 & 7.06 & 7.89 & \bf 9.33\\
\bottomrule
\end{tabular}
}
\caption{Monosemanticity score on different layer and expnsion factor. Higher is better.}
\label{tab:ms}
\end{table}

\noindent\textbf{Limitations.} 
We illustrate our approach on Qwen2-Audio-7B-Instruct~\cite{chu2024qwen2}, focusing on a representative decoder layer under the assumptions of universality and disentanglement~\cite{bricken2023monosemanticity,templeton2024scaling}. While the probing dataset is moderate in size, it reliably surfaces coherent monosemantic concepts. The framework is lightweight, scalable, and readily extendable to larger datasets, deeper layers, and other AudioLLMs.

\section{Conclusion}
We introduced AR\&D, the first mechanistic interpretability pipeline for AudioLLMs, using sparse autoencoders to disentangle polysemantic activations into monosemantic features. Our method discovers representative audio clips, names features via automated captioning, and validates them through human evaluation and steering. Results show that AudioLLMs encode structured, interpretable features, enhancing transparency and control. This work offers an initial step toward enabling trustworthy deployment of AudioLLMs, with future directions including larger models, multilingual audio, and fine-grained paralinguistic features.

\clearpage
\bibliographystyle{IEEEbib}
{\small
\bibliography{main}

@inproceedings{huben2023sparse,
  title={Sparse autoencoders find highly interpretable features in language models},
  author={Huben, Robert and Cunningham, Hoagy and Smith, Logan Riggs and Ewart, Aidan and Sharkey, Lee},
  booktitle={ICLR},
  year={2024}
}

@article{dreyer2025mechanistic,
  title={Mechanistic understanding and validation of large AI models with SemanticLens},
  author={Dreyer, Maximilian and Berend, Jim and Labarta, Tobias and Vielhaben, Johanna and Wiegand, Thomas and Lapuschkin, Sebastian and Samek, Wojciech},
  journal={Nature Machine Intelligence},
  pages={1--14},
  year={2025},
}

@inproceedings{gaoscaling,
  title={Scaling and evaluating sparse autoencoders},
  author={Gao, Leo and la Tour, Tom Dupre and Tillman, Henk and Goh, Gabriel and Troll, Rajan and Radford, Alec and Sutskever, Ilya and Leike, Jan and Wu, Jeffrey},
  booktitle={ICLR},
  year={2025}
}

@inproceedings{elizalde2023clap,
  title={Clap learning audio concepts from natural language supervision},
  author={Elizalde, Benjamin and Deshmukh, Soham and Al Ismail, Mahmoud and Wang, Huaming},
  booktitle={ICASSP},
  year={2023}
}

@article{pach2025sparse,
  title={Sparse autoencoders learn monosemantic features in vision-language models},
  author={Pach, Mateusz and Karthik, Shyamgopal and Bouniot, Quentin and Belongie, Serge and Akata, Zeynep},
  journal={arXiv preprint arXiv:2504.02821},
  year={2025}
}

@article{mei2024wavcaps,
  title={Wavcaps: A chatgpt-assisted weakly-labelled audio captioning dataset for audio-language multimodal research},
  author={Mei, Xinhao and Meng, Chutong and Liu, Haohe and Kong, Qiuqiang and Ko, Tom and Zhao, Chengqi and Plumbley, Mark D and Zou, Yuexian and Wang, Wenwu},
  journal={IEEE/ACM Transactions on Audio, Speech, and Language Processing},
  pages={3339--3354},
  year={2024}
}

@article{busso2008iemocap,
  title={IEMOCAP: Interactive emotional dyadic motion capture database},
  author={Busso, Carlos and Bulut, Murtaza and Lee, Chi-Chun and Kazemzadeh, Abe and Mower, Emily and Kim, Samuel and Chang, Jeannette N and Lee, Sungbok and Narayanan, Shrikanth S},
  journal={Language resources and evaluation},
  volume={42},
  pages={335--359},
  year={2008},
  publisher={Springer}
}

@article{chu2024qwen2,
  title={Qwen2-audio technical report},
  author={Chu, Yunfei and Xu, Jin and Yang, Qian and Wei, Haojie and Wei, Xipin and Guo, Zhifang and Leng, Yichong and Lv, Yuanjun and He, Jinzheng and Lin, Junyang and others},
  journal={arXiv preprint arXiv:2407.10759},
  year={2024}
}

@article{fonseca2021fsd50k,
  title={Fsd50k: an open dataset of human-labeled sound events},
  author={Fonseca, Eduardo and Favory, Xavier and Pons, Jordi and Font, Frederic and Serra, Xavier},
  journal={IEEE/ACM Transactions on Audio, Speech, and Language Processing},
  volume={30},
  pages={829--852},
  year={2021},
  publisher={IEEE}
}

@inproceedings{zhang2024largemultimodalmodelsinterpret,
  title={Large Multi-modal Models Can Interpret Features in Large Multi-modal Models},
  author={Kaichen Zhang and Yifei Shen and Bo Li and Ziwei Liu},
  year={2025},
  booktitle={ICCV},
}

@article{wang2024audiobench,
  title={AudioBench: A Universal Benchmark for Audio Large Language Models},
  author={Wang, Bin and Zou, Xunlong and Lin, Geyu and Sun, Shuo and Liu, Zhuohan and Zhang, Wenyu and Liu, Zhengyuan and Aw, AiTi and Chen, Nancy F},
  journal={NAACL},
  year={2025}
}

@article{kuhn1955hungarian,
  title={The Hungarian method for the assignment problem},
  author={Kuhn, Harold W},
  journal={Naval research logistics quarterly},
  volume={2},
  pages={83--97},
  year={1955},
  publisher={Wiley Online Library}
}

@inproceedings{bert-score,
  title={BERTScore: Evaluating Text Generation with BERT},
  author={Zhang, Tianyi and Kishore, Varsha and Wu, Felix and Weinberger, Kilian Q and Artzi, Yoav},
  booktitle={International Conference on Learning Representations},
  year={2020}
}

@article{elhage2022toy,
  title={Toy models of superposition},
  author={Elhage, Nelson and Hume, Tristan and Olsson, Catherine and Schiefer, Nicholas and Henighan, Tom and Kravec, Shauna and Hatfield-Dodds, Zac and Lasenby, Robert and Drain, Dawn and Chen, Carol and others},
  journal={arXiv preprint arXiv:2209.10652},
  year={2022}
}

@inproceedings{tangsalmonn,
  title={SALMONN: Towards Generic Hearing Abilities for Large Language Models},
  author={Tang, Changli and Yu, Wenyi and Sun, Guangzhi and Chen, Xianzhao and Tan, Tian and Li, Wei and Lu, Lu and MA, Zejun and Zhang, Chao},
  booktitle={ICLR},
    year={2024}
}

@inproceedings{kang2024frozen,
  title={Frozen large language models can perceive paralinguistic aspects of speech},
  author={Kang, Wonjune and Jia, Junteng and Wu, Chunyang and Zhou, Wei and Lakomkin, Egor and Gaur, Yashesh and Sari, Leda and Kim, Suyoun and Li, Ke and Mahadeokar, Jay and others},
  booktitle={INTERSPEECH},
  year={2024}
}

@article{olsson2022context,
  title={In-context learning and induction heads},
  author={Olsson, Catherine and Elhage, Nelson and Nanda, Neel and Joseph, Nicholas and DasSarma, Nova and Henighan, Tom and Mann, Ben and Askell, Amanda and Bai, Yuntao and Chen, Anna and others},
  journal={arXiv preprint arXiv:2209.11895},
  year={2022}
}

@inproceedings{lou2025sae,
  title={SAE-V: Interpreting Multimodal Models for Enhanced Alignment},
  author={Hantao Lou and Changye Li and Jiaming Ji and Yaodong Yang},
  booktitle={ICML},
  year={2025},
}

@article{jiao2024audio,
  title={Audio-visual modelling in a clinical setting},
  author={Jiao, Jianbo and Alsharid, Mohammad and Drukker, Lior and Papageorghiou, Aris T and Zisserman, Andrew and Noble, J Alison},
  journal={Scientific Reports},
  volume={14},
  number={1},
  pages={15569},
  year={2024},
  publisher={Nature Publishing Group UK London}
}

@article{CarreiroMartins2024,
  author    = {Carreiro-Martins, P. and Paixão, P. and Caires, I. and Matias, P. and Gamboa, H. and Soares, F. and Gomez, P. and Sousa, J. and Neuparth, N.},
  title     = {Acoustic and Clinical Data Analysis of Vocal Recordings: Pandemic Insights and Lessons},
  journal   = {Diagnostics (Basel, Switzerland)},
  year      = {2024},
  volume    = {14},
  pages     = {2273}
}

@article{bricken2023monosemanticity,
   title={Towards Monosemanticity: Decomposing Language Models With Dictionary Learning},
   author={Bricken, Trenton and Templeton, Adly and Batson, Joshua and et al.},
   year={2023},
   journal={Transformer Circuits Thread},
}

@inproceedings{kim2024paralinguistics,
  title={Paralinguistics-aware speech-empowered large language models for natural conversation},
  author={Kim, Heeseung and Seo, Soonshin and Jeong, Kyeongseok and Kwon, Ohsung and Kim, Soyoon and Kim, Jungwhan and Lee, Jaehong and Song, Eunwoo and Oh, Myungwoo and Ha, Jung-Woo and others},
  booktitle={NIPS},
  year={2024}
}

@article{templeton2024scaling,
   title={Scaling Monosemanticity: Extracting Interpretable Features from Claude 3 Sonnet},
   author={Templeton, Adly and Conerly, Tom and Marcus, Jonathan and et al.},
   year={2024},
   journal={Transformer Circuits Thread},
}

@book{elad2010sparse,
  title={Sparse and redundant representations: from theory to applications in signal and image processing},
  author={Elad, Michael},
  year={2010},
  publisher={Springer Science \& Business Media}
}

@article{olshausen1996emergence,
  title={Emergence of simple-cell receptive field properties by learning a sparse code for natural images},
  author={Olshausen, Bruno A and Field, David J},
  journal={Nature},
  volume={381},
  pages={607--609},
  year={1996}
}

@misc{SeaLLMs-Audio,
    author = {Chaoqun Liu and Mahani Aljunied and Guizhen Chen and Hou Pong Chan and Weiwen Xu and Yu Rong and Wenxuan Zhang},
    title = {SeaLLMs-Audio: Large Audio-Language Models for Southeast Asia},
    year = {2025},
    publisher = {GitHub},
    journal = {GitHub repository},
    howpublished = {\url{https://github.com/DAMO-NLP-SG/SeaLLMs-Audio}},
}

@article{makhzani2013k,
  title={K-sparse autoencoders},
  author={Makhzani, Alireza and Frey, Brendan},
  journal={arXiv preprint arXiv:1312.5663},
  year={2013}
}
}
\end{document}